\documentclass[sigplan]{acmart}
\renewcommand\footnotetextcopyrightpermission[1]{}
\usepackage{tikz}
\usepackage{amsmath}

\usepackage{filecontents}

\usepackage{textcomp} 

\usepackage{multirow}
\usepackage{makecell}
\usepackage{verbatim}
\usepackage{paralist}
\usepackage{subcaption}
\usepackage{graphicx}
\usepackage{balance}
\usepackage{rotating}
\usepackage{pifont}
\usepackage{tikz}
\usepackage{color}
\usepackage{xcolor}
\usepackage{framed}
\usepackage[para,flushleft,online]{threeparttable}
\usepackage{paralist}
\usepackage{flushend}
\usepackage[labelfont={bf}, textfont={}]{caption}
\usepackage{listings}
\usepackage{soul}
\usepackage{enumitem}

\usepackage{tabularx}
\usepackage{booktabs}
\usetikzlibrary{arrows,calc,shapes,automata,backgrounds,petri,matrix,positioning,fit}

\definecolor{mygreen}{rgb}{0,0.6,0}
\definecolor{mygray}{rgb}{0.5,0.5,0.5}
\definecolor{mymauve}{rgb}{0.58,0,0.82}
\newcommand{\pname}[1]{{{LearnedSSD}}{#1}}
\newcommand{\pdb}[1]{{ConfDB}{#1}}

\newcommand*\circleb[1]{\tikz[baseline=(char.base)]{
            \node[shape=circle,draw,inner sep=0.5pt,fill=black,text=white, scale=0.85] (char) {#1};}}

\acmConference{Technical Report, UIUC}{2021}{}

\begin{document}

%

\date{}

\title{A Learning-based Approach Towards Automated Tuning of SSD Configurations}

%
\author{Daixuan Li and Jian Huang}
\affiliation{Systems Platform Research Group\\ University of Illinois at Urbana-Champaign}

\begin{abstract}
Thanks to the mature
manufacturing techniques, solid-state drives (SSDs) are highly customizable for applications today, 
which brings opportunities to further improve their storage performance and resource utilization. However, 
the SSD efficiency is usually determined by many hardware parameters, making it hard for developers 
to manually tune them and determine the optimal SSD configurations.  

In this paper, we present an automated learning-based framework, named \pname{}, that utilizes 
both supervised and unsupervised machine learning (ML) techniques to drive the tuning of hardware configurations 
for SSDs. 
\pname{} automatically extracts the unique access patterns
of a new workload using its block I/O traces, maps the workload to previously workloads for
utilizing the learned experiences, and recommends an optimal SSD configuration based on the validated  
storage performance. \pname{} accelerates the development of new SSD devices by automating the hardware 
parameter configurations and reducing the manual efforts. 
We develop \pname{} with simple yet effective learning algorithms that can run efficiently on multi-core CPUs. 
Given a target storage workload, our evaluation shows that \pname{} can always deliver an optimal SSD 
configuration for the target workload, and this configuration will not hurt the performance of non-target workloads. 

\end{abstract}

\maketitle
\renewcommand{\shorttitle}{A Learning-Based Approach Towards Automated Tuning of SSD Configurations}
%

\section{Introduction}
\label{sec:intro}

Flash-based solid-state drive (SSDs) have become the backbone of modern storage infrastructures in various 
computing platforms, as they offer orders-of-magnitude better performance than hardware-disk drives (HDDs),  
while their cost is approaching to that of HDDs~\cite{willow,flashblox,DesignTradeoff,decibel:nsdi2017, 
mittos:sosp2017, sdf:asplos2014,storpool}. Thanks to the development of manufacturing and shrinking 
process technology~\cite{www:ssdproduct}, the industry has been able to rapidly produce SSD 
devices with different hardware configurations. 

Although SSD devices are becoming highly customizable to meet the ever-increasing demands on storage 
performance and capacity for new applications~\cite{sdf:asplos2014, lightnvm}, 
identifying optimal device configurations is on the critical path
of SSD development.
This is because the SSD hardware configurations are usually determined by the requirements from
applications and customers~\cite{ssdmade, ssddesign}, and these configurations involve many components in the
storage controller, such as flash chip specifications, chip layout, block/page sizes, device buffer 
sizes, and so on. 
In order to deliver optimal performance for applications in a new generation of storage
device development, storage vendors usually use
typical application workloads as their benchmarks to aid them to determine the device configurations.
However, an SSD device has hundreds of parameters in its configurations,
and these parameters
usually have dependencies (i.e., the update of one parameter may affect other parameters),
making it hard for hardware engineers to tune the device configurations and identify the optimal ones
in a short time. This significantly hurts the productivity of new SSD device development~\cite{ssdmade}.


Furthermore, there is an increasing demand for customized storage devices from various computing
platforms and applications. This is for two reasons.
First, computing platforms always wish to deploy the best-fit storage devices
for their workloads, such that they can achieve the maximum performance.
This is especially true for cloud platforms that require highly customized SSDs to support their
cloud services, such as Database-as-a-Service~\cite{www:dbservice} and web services~\cite{www:webservice}.
As applications such as cloud storage services are evolving quickly, we need to revolutionize the 
configuration tuning procedure to shorten the lifecycle of producing new SSDs.   

Second, our study shows that storage workloads can be categorized with learning algorithms,
which provides the evidence that it is feasible to customize storage devices for a specific workload 
type (see Figure~\ref{fig:cluster}), especially considering the SSD manufacturing techniques become 
mature today. 
However, there is still a long-standing gap between application demands
and SSD device configurations. And our community lacks a framework that can instantly
transfer application demands into device configurations of SSDs.

In this paper, we develop an automated framework named \pname{}, which exploits both supervised and 
unsupervised machine learning (ML) techniques to drive the hardware configurations for new SSDs. 
Given a storage workload, \pname{} will recommend an optimal SSD configuration that delivers  
optimized storage performance. It leverages the linear 
regression techniques to expose the device configurations that have the strongest correlation to the 
storage performance. To present reasonable device configurations, we formulate different types of 
hardware parameters in the SSD, transfer them into the vectors in the ML model, and utilize learning techniques 
to explore the optimization space and identify the near-best options, with specified 
constraints such as SSD capacity, interfaces (NVMe or SATA), and flash memory types. 

To reduce the execution time of learning an optimal SSD configuration while ensuring the learning 
accuracy, we develop pruning algorithms to identify the most important hardware parameters in SSDs. 
\pname{} also maintains a configuration database named \pdb{} that stores the learned workloads and 
SSD configurations. For a new workload, \pname{} will extract its features and compare 
them with the records in the \pdb{} using similarity comparison networks. 
If \pname{} identifies a similar workload in its \pdb{}, it 
will recommend the corresponding SSD configuration directly, such that we can utilize the previously  
learned experience. Otherwise, \pname{} will learn a new SSD configuration for the workload, and 
add them into its \pdb{} for future references. As many storage workloads share similar data access 
patterns and can be categorized into a general type (see Figure~\ref{fig:cluster}), 
\pname{} can assist developers to identify the most critical parameters for a type of storage workloads, 
and recommend an optimal SSD configuration. 

Our study with \pname{} leads to interesting insights. We summarize them into learning 
rules that can aid developers to prioritize their optimization strategies when producing new 
SSDs. For instance, (1) not all SSD parameters are equal, 
the layout arrangement of flash chips is an important factor for storage performance; 
(2) with different targets and configuration constraints, the tuning prcedure of device configurations are different, 
and for each parameter, its correlation with SSD performance is also different; 
(3) not all parameters are sensitive to storage performance, and some of them can be 
configured as the same as commodity SSDs today. 

To evaluate the efficiency of \pname{}, we implemented our proposed techniques using PyTorch~\cite{www:pytorch}, 
the \textit{scikit-learn} tool~\cite{www:scikit}, and a production-level SSD simulator 
MQSim~\cite{mqsim:fast2018}. 
We perform experiments with a variety of storage traces. Our experimental results show that \pname{} delivers 
an SSD configuration that can achieve 1.28--34.61$\times$ performance improvement for a target workload, 
without hurting the performance of non-target workloads, compared to the configurations specified by the released 
commodity SSDs. We also show that \pname{} can learn a new configuration in 37.3 seconds, and finalize an optimal 
configuration in 121 iterations on average for a workload, with a multi-core server processor. 
Overall, we make the following contributions:  

\begin{itemize}[leftmargin=*]

\item We present the first study of SSD hardware parameters and popular storage workloads with 
	learning in mind, and demonstrate the feasibility of applying the learning-based approach for 
	identifying optimal SSD specifications. 

\vspace{0.85ex}
\item We formulate the tuning problem of SSD device configurations using a learning-based approach, and 
develop an automated learning framework that can efficiently recommend optimal 
SSD configurations for different workloads.  

\vspace{0.85ex}
\item We summarize a set of learning rules that can facilitate the hardware configurations and 
development of new SSDs, based on our study with \pname{}. 

\vspace{0.85ex}
\item We study the efficiency of \pname{} and show its benefits in comparison with released 
commodity SSD settings. 

\end{itemize}

\section{Background and Motivation}
\label{sec:motivation}


SSD has proven to be a revolutionary storage technology. It performs much faster   
than hard-disk drives (HDDs), while its price is reaching to that of HDDs.
The rapidly shrinking process and manufacturing technologies have accelerated the SSD device 
development and enabled their widespread adoption 
in a variety of computing platforms, such as data centers~\cite{sdf:asplos2014,www:opencompute,storpool,www:denali}.

\begin{figure}[t]
\includegraphics[width=0.9\linewidth]{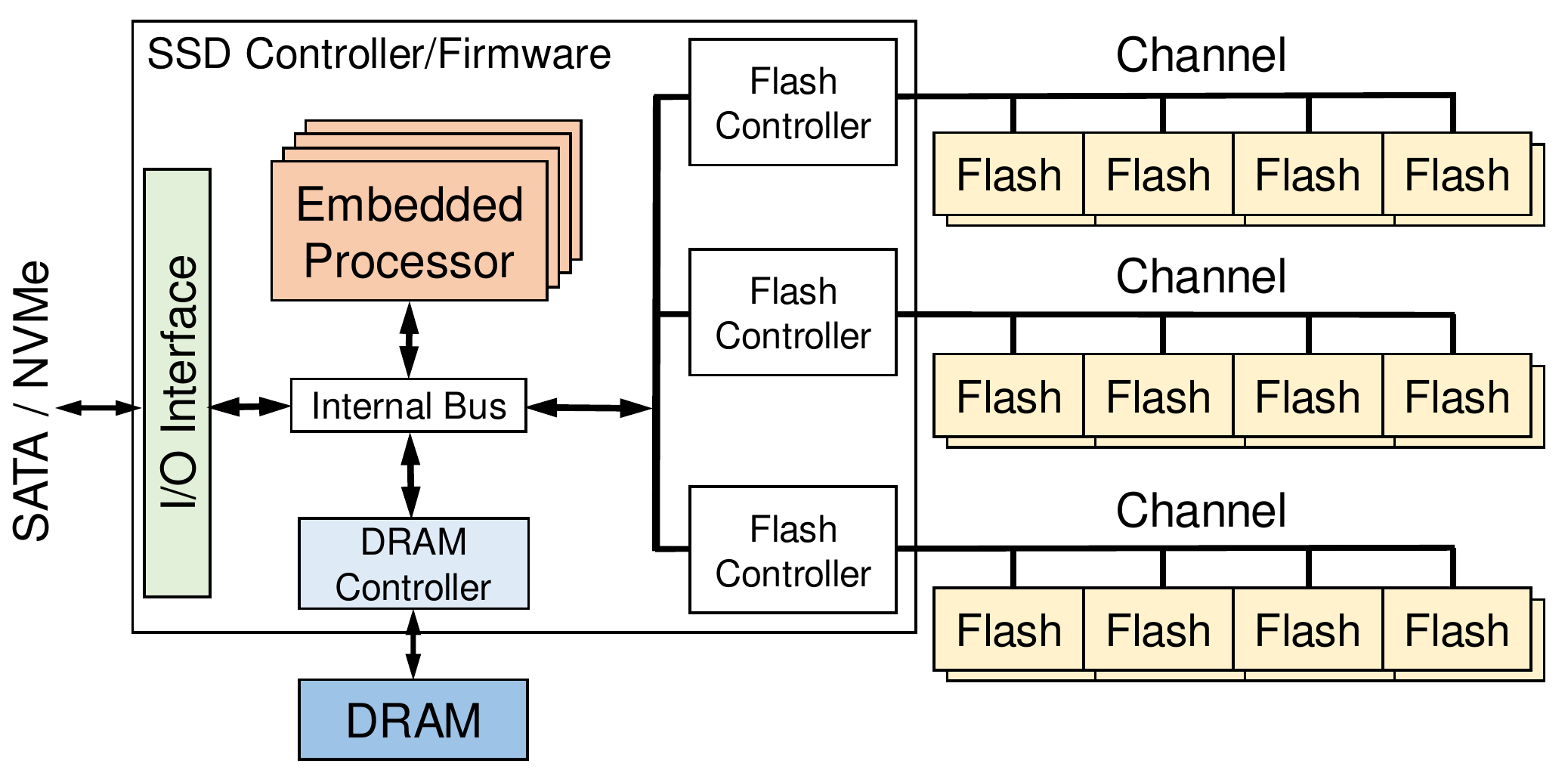}
\vspace{-2ex}
\caption{Internal architecture of flash-based SSDs.}
\label{fig:ssd-arch}
\vspace{-2ex}
\end{figure}

\subsection{SSD Architecture}
We present the internal system architecture of a typical SSD in Figure~\ref{fig:ssd-arch}.
An SSD consists of five major components: a set of flash memory packages,
an SSD controller having embedded processors like ARM, off-chip DRAM (SSD DRAM), 
flash controllers, and the I/O interface that includes SATA and NVMe protocols~\cite{chen2011essential, 
dftl, summarizer}. 
The flash packages are organized in a hierarchical manner. Their organization 
not only determines the storage capacity but also affects the storage performance. Each SSD has multiple 
channels where each channel can receive and process read/write commands independently. 
Each channel is shared by multiple flash packages. Each package has multiple flash chips. 
Within each chip, there are multiple planes. Each plane includes multiple flash blocks, 
and each block consists of multiple flash pages. And the page size varies in different SSDs. 

In order to manage the flash memory, the SSD controller usually implements the 
Flash Translation Layer (FTL) in its firmware. The FTL was developed for 
taking care of the intrinsic properties of SSDs. When a free flash page is written once, 
that page is no longer available for future writes until that page is erased.
However, erase operation is performed only at a block  
granularity. To remove the expensive erase operations from the critical path, 
writes are issued to free pages that have been erased, which is also called out-of-place update. 
The SSD controller will perform Garbage Collection (GC) later to clean the stale data.
As each flash block has limited endurance, it is important for blocks to age uniformly (i.e., wear leveling).
Modern SSD controllers employ out-of-place write, GC, and wear leveling to overcome these aforementioned 
shortcomings and maintain indirections for handling the address mapping in their FTL.

\subsection{SSD Manufacturing and Parameter Tuning}
\label{subsec:manufacturing}
According to the interviews with SSD product managers~\cite{www:ssdmade} and our discussions  
with SSD vendors, finalizing the SSD hardware specifications or parameters is on the critical path 
in the SSD design. These specifications are usually determined by the requirements from applications 
and customers. And they involve the components of the SSD controller as shown in Figure~\ref{fig:ssd-arch}.  
Without these specifications, the SSD development cannot proceed to the manufacturing stage. 
With the confirmation from SSD vendors, there are more than a hundred tunable parameters in a typical SSD. 

To finalize the SSD specifications, a straightforward approach is to test and profile 
application workloads with different hardware configurations.  
However, this is not scalable as we target different application workloads. 
Given an application workload, it is challenging  
for developers to test all the combinations of the device parameters. And likewise, give a new SSD 
specification, 
it requires significant manual effort to quantify the effectiveness of the selected specifications. 
In this work, we use the learning-based approach to automate the SSD hardware configurations.

\subsection{Software-Defined Solid-State Drive}
With the increasing demands on storage performance from applications, we have seen a trend that 
modern storage systems are embracing software-defined hardware techniques~\cite{www:opencompute, 
www:denali}. This allows upper-level applications to 
achieve maximum performance benefits and resource efficiency with customized storage devices. 
For instance, the recent development of software-defined SSDs~\cite{lightnvm, flashblox:fast17} 
enables platform operators to customize the number of flash channels and chips in an SSD, 
with the cooperation with SSD vendors. 

\begin{figure}[t]
	\centering
\includegraphics[scale=0.65]{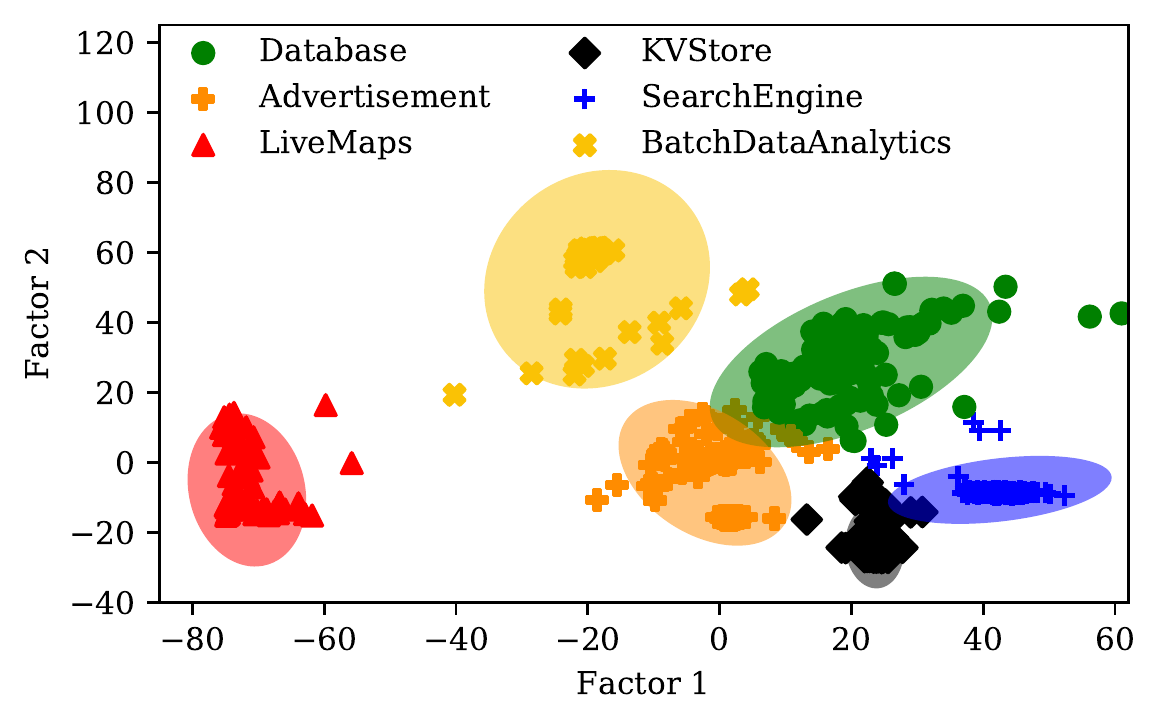}
\vspace{-3ex} 
\caption{A clustering of popular storage workloads.} 
\label{fig:cluster} 
\end{figure}

This is especially true for both public and private cloud platforms that require highly customized 
SSDs to support their cloud services, such as database-as-a-service~\cite{www:dbservice}, 
web services~\cite{www:webservice}, web search~\cite{flashblox:fast17}, and batch data analytics~\cite{mapreduce}. 
For these applications, their workloads can be highly classified. 
For instance, we use our proposed learning-based workload characterization approach (see the detailed 
discussion in $\S$\ref{subsec:clustering}) to study the storage traces from a set of 
popular application workloads (see Table~\ref{tab:workloads}). Our experiments demonstrate that 
each workload type has its unique characteristics, and I/O traces from the same workload type have 
similarities in their data access pattern, as shown in Figure~\ref{fig:cluster}. 
This provides the evidence showing that it is feasible to customize storage devices for a specific 
category of applications. 
Unfortunately, our community lacks a framework that can efficiently 
transfer application demands and characteristics into the hardware configurations of SSDs. 

\subsection{Learning-Based Parameter Tuning}

We have seen a disruptive advancement of machine learning (ML) techniques over the past decade. 
In general, we can categorize machine learning techniques into two types: supervised learning and 
unsupervised learning. As for supervised learning, it learns a set of rules with labeled datasets and
then generalizes these rules to make predications for new inputs. Typical example algorithms include
decision trees~\cite{dtree}, support vector machines~\cite{svm},
Bayesian networks~\cite{bnetwork}, and artificial neural networks~\cite{nnetwork}.
Unlike supervised learning, the unsupervised learning can identify unknown patterns based on
unlabeled datasets, such as k-means clustering and principal component analysis (PCA)~\cite{chris:icml2004}.

Recent studies~\cite{aichip:nature, firm:osdi2020, peakprediction:eurosys2021,  
microservices:euromlsys2021, mldatabase:dataengr2021, ottertune:sigmod2017, 
akgun:hotstorage2021, yanqi:asplos2021, yu:asplos2021}
showed that the learning-based method is a promising approach to solve system optimization problems. 
However, none of them investigated their applications in SSD development, and it is unclear 
how can we utilize learning-based approaches to overcome the challenges of tuning hardware 
specifications of SSDs. 
In this work, we will use both supervised learning and unsupervised learning to develop \pname{}.

\section{Design and Implementation}
\label{sec:design}
In this section, we first discuss the goals of \pname{}. 
And we will present the system architecture of \pname{}, and its 
core components respectively.

\begin{figure*}[t] 
\includegraphics[width=0.78\textwidth]{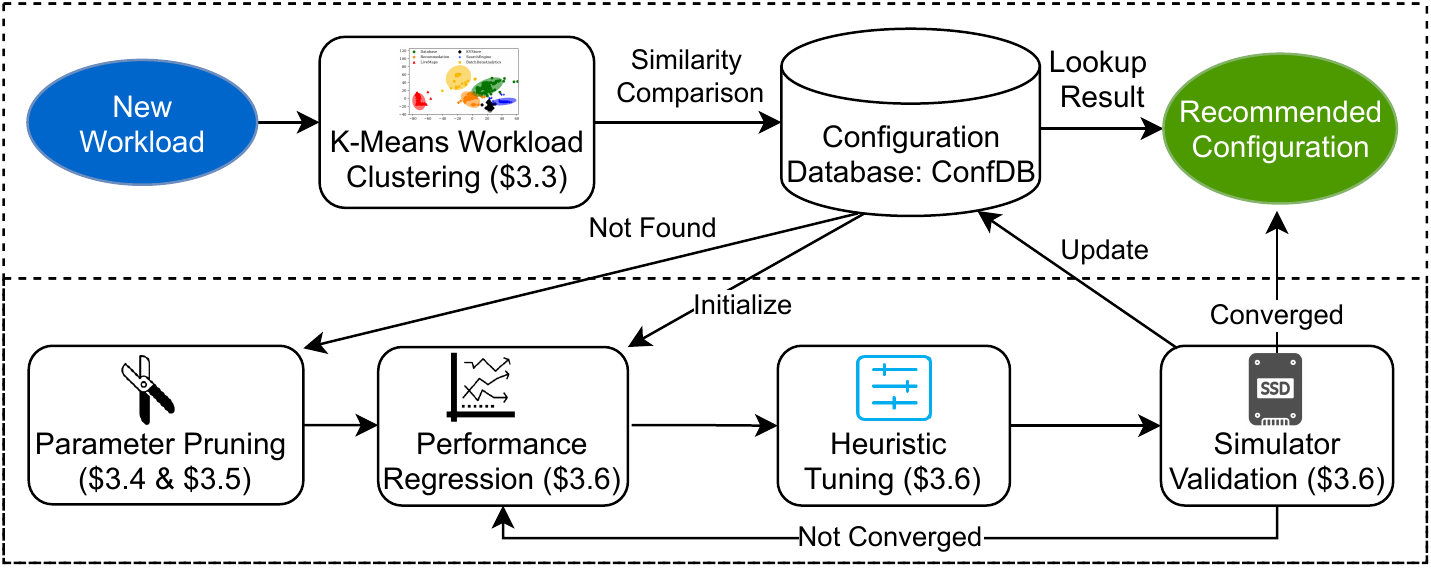}
\vspace{-1ex} 
	\caption{System overview of \pname{}:\ \ \pname{} first learns new workload features with \textbf{a learning-based workload clustering ($\S$\ref{subsec:clustering})}. 
	If it is similar to workloads in existing clusters in the configuration database \pdb{}, \pname{} will recommend an optimal configuration stored in the database. 
	If not, \pname{} will first conduct \textbf{parameter pruning ($\S$\ref{subsec:parameter} and $\S$\ref{subsec:pruning})} to identify performance-critical 
	parameters of SSDs. After that, \pname{} will conduct automated tuning which consists of three modules: \textbf{performance regression} with Gaussian process, 
	\textbf{heuristic tuning, and performance validation ($\S$\ref{subsec:autotuning}).}
	} 
\label{fig:overview} 
\end{figure*}

\subsection{Design Goals}
\label{subsec:goal}
The high-level goal of \pname{} is to enable the automated tuning of hardware configurations of SSDs for 
a specific application workload with learning techniques. Specifically, \pname{} will achieve the 
following goals: 

\begin{itemize}[leftmargin=*]

\vspace{0.8ex}
\item It can generate an optimal SSD configuration for a target workload, 
while this hardware configuration has minimal negative impact on other workloads. 

\vspace{0.8ex}
\item It can identify an optimal SSD configuration in a short time, without 
introducing much extra computation overheads.

\vspace{0.8ex}
\item It can scale to support diverse target workloads as well as different device constraints.  

\end{itemize}

In the following, we will discuss how \pname{} achieves these goals and overcomes the challenges in both 
systems building and configuration learning procedure.

\subsection{System Overview}
\label{subsec:overview}

We develop an automated framework named \pname{}, which
exploits both supervised and unsupervised learning techniques to drive the specifications for new SSD devices.
Given a storage workload and design constraints, \pname{} will recommend an optimal SSD configuration that can 
deliver optimized storage performance shortly.

To be specific, we leverage linear regression techniques to expose the device specifications that have the strongest 
correlation with the storage efficiency. To present reasonable device specifications, we formulate different types 
of hardware parameters in the SSD, transfer them into the vectors of the learning models, and 
learn the optimal options with specified constraints, such as storage capacity and SSD interfaces (NVMe or SATA).
To reduce the execution time of learning an optimal SSD configuration while ensuring the learning
accuracy, we develop pruning algorithms to identify the most important hardware parameters in SSDs.
We will also maintain a configuration database named \pdb{} to store the learned workloads 
and the corresponding SSD specifications. Therefore, we can utilize the learned experiences for new workloads 
in \pname{}.

For a new workload, \pname{} will extract its features and compare
them with the records in the configuration database \pdb{}, as shown in Figure~\ref{fig:overview}.
If \pname{} identifies similar workloads, 
it will either recommend a known optimal specification, or retune the SSD configurations with the recorded SSD specifications, 
such that we can utilize previous learned experiences. Otherwise, \pname{} will learn new SSD specifications and
add the learned configurations into \pdb{} for future references.
As many storage workloads 
can be categorized into a general type (see Figure~\ref{fig:cluster}),
\pname{} can also assist developers to identify the most critical parameters for a type of storage workloads. 
It is worth noting that \pname{} will ensure the recommended SSD
specifications will not hurt the storage performance for other generic workloads.

\subsection{Learning-based Workload Clustering}
\label{subsec:clustering}
Unlike traditional ways of using the read/write ratio,
and I/O patterns (e.g., sequential and random read/write) to categorize workloads, which cannot capture the
whole picture of workload characteristics, we develop a learning-based clustering approach based on block I/O traces.
We chose block I/O traces to learn the characteristics of storage workloads, because this approach does 
not have system dependencies and not require application semantics. 

To develop the learning-based workload clustering, we first partition each I/O trace into small windows. 
According to our study of diverse workloads, we use 3,000 trace entries in each window by default.  
This is because fewer entries may lose the unique data access patterns of the trace, and more entries would generate 
less valid data points in a cluster, they both could hurt the accuracy of the workload clustering. 
The trace information used to conduct the workload characterization include 
I/O timestamp, I/O size, device number, block address, and operation types. 
We convert each window of the I/O traces into a data point, and use Principal Component Analysis~\cite{pca} to transfer
the data points into two dimensions. After that, we use k-means to cluster these data points. 

After we cluster all the data points of a new workload, we will calculate the distance between 
the center of the examined data points and the center of an existing cluster. If the distance is below a threshold, we make 
the conclusion that the majority of the examined data points fall into the existing workload cluster. In other words, 
the new workload belongs to this cluster. 
If \pname{} cannot identify a similar cluster, \pname{} will create a new cluster for the new workload.
Our experimental results, as demonstrated in Figure~\ref{fig:cluster}, show that
our learning-based workload clustering can successfully identify a cluster of storage workloads that belong to the same or 
similar workload type. 

To further verify the effectiveness of the learning-based approach for workload clustering, we divide the same workload into the 
training and validation datasets, and evaluate the clustering results for each workload as listed in Table~\ref{tab:workloads}. 
We observe that 95\% of their data points fall into the same workload cluster on average. 
This provides the evidence showing that our proposed learning-based approach is sufficient to identify an appropriate cluster for new workloads. 
It also drives the \pname{} design by offering the insight that it is feasible to develop an SSD which can deliver 
optimized performance for a category of application workloads.

\subsection{Transfer SSD Specifications into ML Parameters}
\label{subsec:parameter}
We now discuss how we can transfer the tuning problem of SSD specifications for a workload category into a ML problem, such that we can 
automate the tuning procedure with high accuracy. To address this problem, we first need to formulate the SSD specifications into ML parameters. 
However, this is not easy, since we have to ensure the parameter formulation does not lose the meaningful semantic of a specific SSD hardware parameter, 
and the ML parameters should be able to represent the characteristics of different hardware specifications and their correlations.  

To model the SSD specifications via ML parameters, we formulate them into three major parts in the ML model: (1) the performance metrics used as 
the optimization targets for SSDs; (2) SSD hardware configurations that can be vectorized as parameters in a ML model; 
and (3) the configuration constraints (e.g., the SSD capacity) that bound the optimization space of the ML model. 
We describe each of them as follows.  

\paragraph{Performance metrics used in the ML model}
As for storage performance, \pname{} focuses on the storage latency and throughput. To quantify whether a SSD configuration delivers 
optimized performance or not, we use reference performance as the baseline (e.g., the latency and throughput obtained from a commercial SSD's configurations), 
and the relative performance improvements as the evaluation metrics. We set the performance optimization goal as follows:  

\begin{equation}
\begin{aligned}
Goal(conf) = &(1-\alpha) \times log(\frac{Latency_{conf}}{Latency_{refer}})  \\ &- \alpha \times log(\frac{Throughput_{conf}}{Throughput_{refer}})
\end{aligned}
\label{equa:goal}
\end{equation} 

where $\alpha$ is a tunable coefficient factor for balancing the latency and throughput at a proper scale. We set $\alpha = 0.9$ by default in \pname{}, based on our study of different coefficients.
In our evaluation (see $\S$\ref{subsec:robustness}), we will examine the impact of the $\alpha$ on the learning efficiency. As discussed, the $Latency_{ref}$ and $Throughput_{ref}$ will be given as the 
reference performance in Formula~\ref{equa:goal} in the ML model.  


\paragraph{SSD hardware specifications as ML parameters}
To represent SSD hardware specifications in the ML models, we transfer them into four types of parameters and use different ways to set their values. 
They include \textit{continuous}, \textit{discrete}, \textit{boolean}, and \textit{categorical} parameters. 

\begin{itemize}[leftmargin=*]
    
	\item Continuous parameter: typical examples of continuous parameter include over-provisioning ratio for GC, and the number of flash channels. 
		To set the value of this type of parameters as we run the ML model, we identify a range of possible values it could take in advance, and 
		divide the range uniformly into \emph{N} small pieces. Therefore, \pname{} can take \emph{N} endpoints as the possible values. 
		For each continuous parameter, we set the range to cover all common values in commodity SSDs for ensuring the learned SSD specifications 
		are practical. 
    
	   \vspace{1ex}
    \item Discrete parameter: typical examples of discrete parameter include the SSD DRAM capacity, I/O queue depth, and page size. We select 
	    all their possible values and store them in a list. Therefore, we can use the list index in the vector of the ML model. 
	    Their possible values also cover all the common values. Each discrete parameter follows different rules as \pname{} sets the value at runtime. 
		For instance, the SSD DRAM capacity will take the power of 2 as we increase it; and there are only five PCIe bandwidth settings, 
		according to the PCIe protocol.
    
	   \vspace{1ex}
    \item Boolean parameter: we use the boolean parameters in the ML model to indicate whether a function or feature (e.g., statistic wear leveling, 
	    and greedy GC) will be enabled in the SSD or not. With the 0-1 boolean parameter, 0/1 means function enabled/disabled respectively.
    
	   \vspace{1ex}
    \item Categorical parameter: As for the categorical parameter, we convert it to the dummy variable~\cite{dummyvariable}. 
	    For example, there are 16 possible values for the plane allocation scheme, we create a list with the length of 16. 
	    When \pname{} selects one scheme, it will set the value of the corresponding index of the list to 1, and others to 0. 
    
\end{itemize}


\begin{figure*}[htb] 
\includegraphics[width=0.98\textwidth]{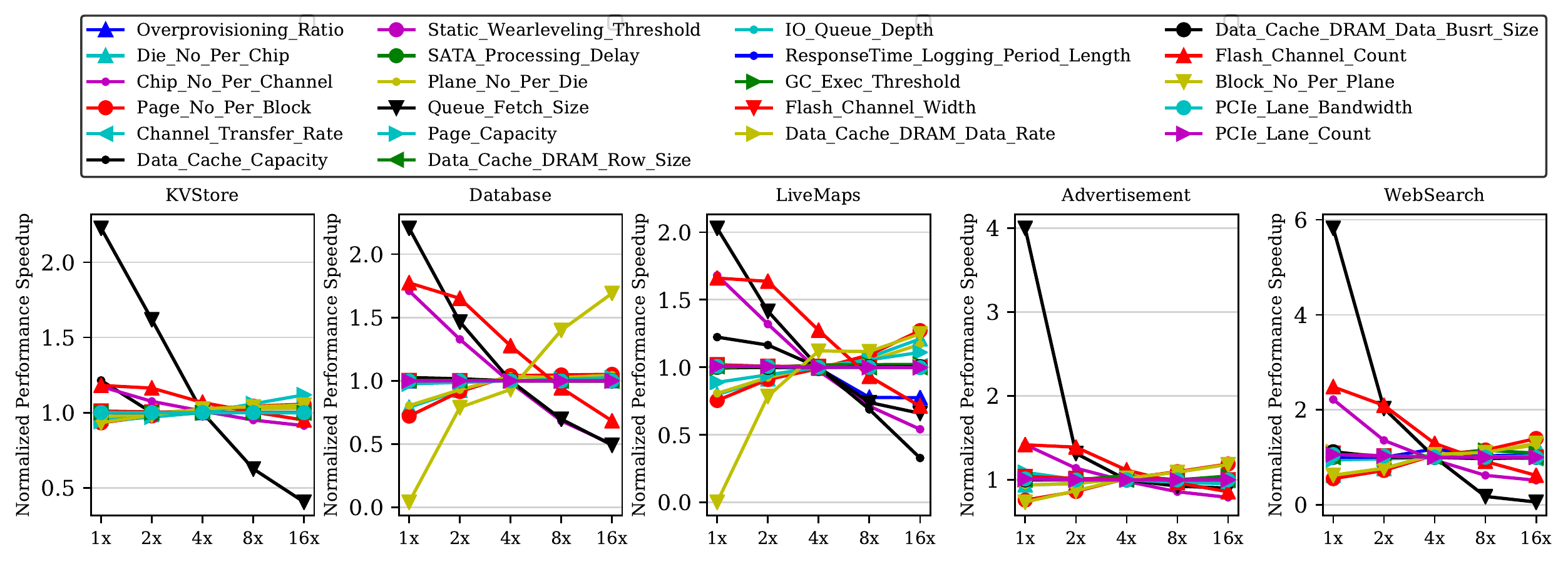}
\vspace{-2ex} 
\caption{A study of coarse-grained parameter pruning for different storage workloads.} 
\label{fig:coarsed_pruning} 
\end{figure*}

\paragraph{Configuration constraints} 
\pname{} allows users to specify the configuration constraints for its configuration tuning. Typical examples include 
the SSD storage capacity and the interface (e.g., NVMe or SATA) supported by the SSD for interacting with the host machine. 
When \pname{} sets different values for 
its ML parameters, it will simply abandon those configurations that violate the specified constraints. And 
\pname{} will always check the constraints when it learns a new configuration. Note that 
\pname{} mainly works on the tunable parameters in SSD specifications. For the lower-level circuit-relevant specifications 
that are strictly limited by the hardware, \pname{} does not cover them in its tuning model.  


\subsection{Learning-based Parameter Pruning}
\label{subsec:pruning}
After we transfer the SSD specifications into ML parameters, we can start to train the model. 
However, modern SSDs usually have hundreds of hardware specifications or parameters. 
Although ML models today can handle a large set of parameters, it is still desirable to develop 
efficient and lightweight models for reducing both training and learning time as well as 
saving computation cycles. 
For example, we develop a model with 64 SSD parameters, it takes 30.7 hours to converge the model 
on a modern multi-core server (see the experimental setup in $\S$\ref{subsec:setup}). 
Moreover, we find that not all SSD parameters are strongly correlated to the storage performance, 
and it is not needed to include these insensitive parameters in the learning model.  
To this end, we propose a parameter pruning approach to identify the impactful parameters 
that could affect the storage performance of SSDs. 
However, we have to overcome two major challenges within the parameter pruning. 
First, we need an accurate measurement method to examine 
the importance of a parameter. This is challenging as SSD parameters usually have dependencies. As we tune a single parameter while keeping 
the values of other parameters fixed, this may violate the configuration constraints. 
For example, increasing the number of flash channels could violate the constraint of the SSD capacity. 
On the other hand, as we tune a single parameter while updating the values of other parameters accordingly for meeting 
the configuration constraints, we cannot accurately determine which parameters affect the storage performance significantly. 
Second, removing some of the SSD parameters may hurt the overall accuracy of the learning model. And it is challenging to quantify 
how each SSD parameter could affect the learning accuracy. 
To address these challenges, we conduct the parameter pruning procedure with two stages.

\paragraph{Coarse-grained parameter pruning.}
We first adopt a coarse-grained pruning method that adjusts the values of continuous and discrete numerical parameters with large stride length. 
But we ensure the values of these parameters are configured in a reasonable range, and still satisfy the configuration constraints. 
At this stage, we eliminate the parameters that do not have much impact on the storage performance, no matter how we change their values. 
As shown in Figure~\ref{fig:coarsed_pruning}, we increase the values of the 23 numerical parameters of SSDs from their baseline setting to 16$\times$, and 
measure the storage performance with different workloads. We observe that some parameters do not affect the storage performance significantly (those 
flat lines in Figure~\ref{fig:coarsed_pruning}), we call them as insensitive parameters in this paper. 
We also find that these insensitive parameters will be different for different storage workload types, therefore, \pname{} will conduct the coarse-grained 
parameter pruning for each workload type and identify the corresponding insensitive parameters. 
In general, we identify 9 insensitive parameters, such as \emph{Page\_Metadata\_Size} and \emph{SATA\_Processing\_Delay}, 
according to our study (see Figure~\ref{fig:coarsed_pruning}). Note that these insensitive parameters would be updated for a new workload type. 

\begin{figure*}[ht] 
\includegraphics[width=0.98\textwidth]{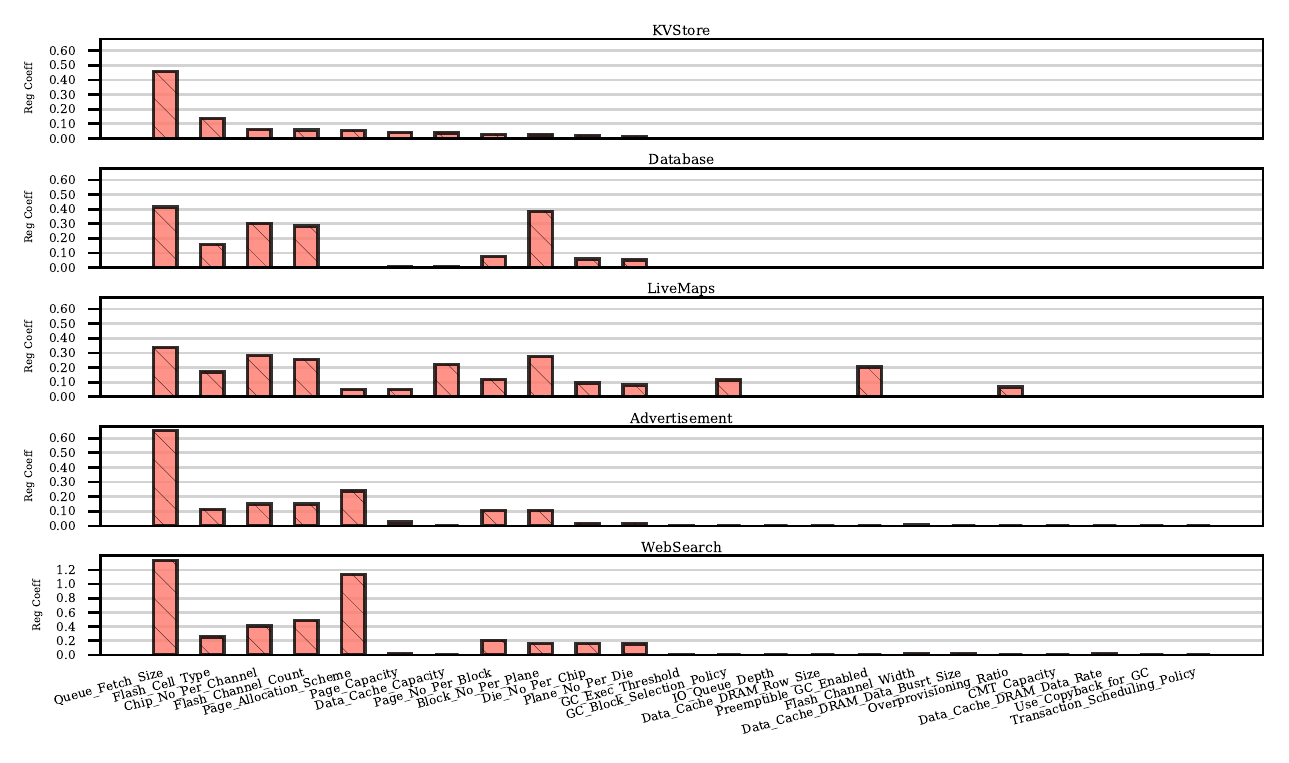}
\vspace{-5ex} 
\caption{A study of fine-grained parameter pruning with linear regression for different storage workloads.} 
\label{fig:fine_grained} 
\end{figure*}

\paragraph{Fine-grained parameter pruning.}
After eliminating the insensitive parameters with the coarse-grained pruning, we continue the parameter pruning with a fine-grained approach. It employs the linear regression 
technique LASSO~\cite{lasso} to identify the linear correlations between the SSD parameters and performance. Following the discussion in $\S$\ref{subsec:parameter}, 
we set a regression space by maintaining the SSD capacity constraint, as we vary the values of SSD parameters. Since the SSD capacity is mainly determined by the parameters 
related to the chip layout, such as flash page size, the number of flash channels, and the flash block size, we first set the values for these parameters. After that, we 
vary the values of other parameters, and measure the regression coefficient for each SSD parameter. A higher regression coefficient score of a parameter means it has a 
closer correlation with the SSD performance. Based on the reported coefficient scores, we abandon 
the parameter whose score is below a threshold (0.01 by default in \pname{}). 
Therefore, we can focus on the parameter tuning for the important ones. 
As shown in Figure~\ref{fig:fine_grained}, we remove the insensitive parameters identified by the coarse-grained parameter tuning, 
and also include the boolean and categorical parameters in the fine-grained parameter tuning. Given the threshold of 0.01 for the regression coefficient score, 
we can eliminate more insensitive parameters for different workloads. 


\paragraph{Observations.}
The learning-based parameter pruning of \pname{} can not only help us to eliminate the insensitive parameters, 
but also offer interesting insights that would benefit SSD development. Specifically, we observe that: (1) 
Different workloads have different parameter sensitivity. For example, the performance of latency-critical workloads like 
Advertisement and WebSearch is not sensitive to the page size and over-provisioning ratio of SSDs, as they are read intensive. 
In contrast, the I/O-intensive workloads such as key-value stores and LiveMaps are sensitive to the flash page size. 
(2) Not all parameters have linear correlation with SSD performance, which generates difficulties for manual tuning, and further 
motivates us to utilize ML techniques to pinpoint the optimal SSD specifications. 


\subsection{Automated Tuning of SSD Configurations}
\label{subsec:autotuning}

\begin{figure*}[!h] 
\includegraphics[width=0.75\textwidth]{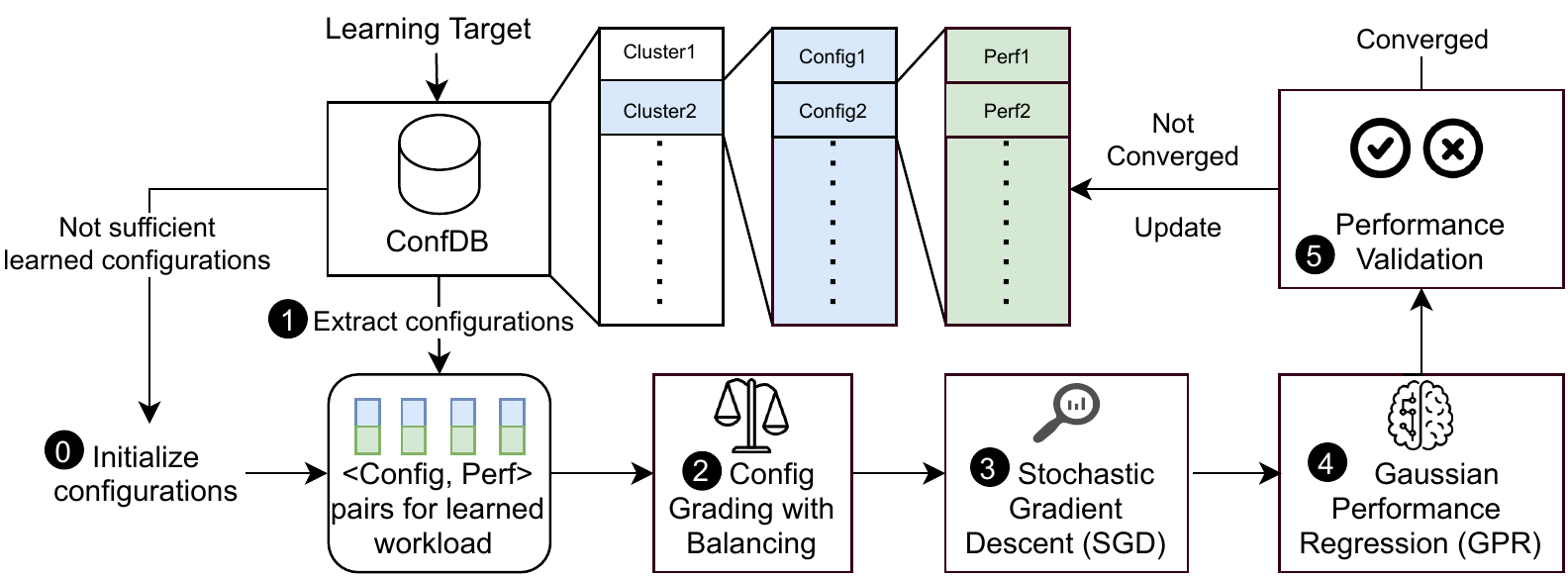}
\vspace{-1ex} 
\caption{The automated learning workflow of \pname{}.} 
\label{fig:overview_workflow} 
\end{figure*}

After the SSD parameter pruning, we now develop the ML model to learn the optimal SSD specifications for various workloads. 
We present the system workflow of \pname{} in Figure~\ref{fig:overview_workflow}. Given a workload, \pname{} will first use the configurations 
stored in the \pdb{} as the initial configuration set, and leverage both Gaussian process regression (GPR) and discrete stochastic gradient descent (SGD) 
algorithms to learn different configurations. For each learned configuration, \pname{} will use a cycle-accurate SSD simulator to validate its 
performance until the model converges (i.e., the optimal SSD configuration is identified). In the following, we discuss each step of the workflow in details.  


\paragraph{Identify the initial configuration set for a new workload.}
For a new workload, \pname{} will use the learning-based workload clustering as discussed in 
$\S$\ref{subsec:clustering} to cluster the workload, and look up the learned configurations for the 
corresponding workload cluster in \pdb{} (\circleb{1}). \pname{} will use these configurations and their 
delivered performance to initialize the ML model. However, if there are insufficient configurations 
in \pdb{} (e.g., the \pdb{} is empty), \pname{} will use a configuration from existing commodity SSDs 
and its measured performance to initialize the model (\circleb{0}).


\paragraph{Quantify the configurations with a unified grading mechanism.}
\pname{} will start the configuration tuning procedure based on the initial configurations. 
In order to check the effectiveness of these configurations, \pname{} develops a grading mechanism (\circleb{2}) 
with the goal of unifying different performance metrics (see $\S$\ref{subsec:parameter}).
To achieve the maximal optimizations for both data access latency and throughput, \pname{} uses the 
Formula~\ref{equa:goal} as the goal. To ensure the learned configuration for a target workload does not 
hurt the performance of other workloads, \pname{} introduces a new factor $\beta$ named \textit{penalty balance} 
in its grading. Therefore, we define the performance grade for a workload as follows:


\begin{equation}
\begin{aligned}
Grade(conf) &= (1-\beta) \times Goal(conf) \\
&+ (\beta) \times \frac{\sum_{conf' \in non-target} Goal(conf')}{NumClusters - 1}
	\label{equa:grade}
\end{aligned}
\end{equation}

where \textit{Goal (conf)} is defined in Formula~\ref{equa:goal}, and $\beta = 0.9$ by default based on 
our study (see Figure~\ref{fig:penalty_balance} in our evaluation).


\paragraph{Search optimal configurations with the stochastic gradient descent technique}
With the initial SSD configurations and their grades, \pname{} will use SGD to search an optimal configuration (\circleb{3}). 
Specifically, \pname{} first identifies the top three best configurations (i.e., the configurations whose grades rank at the top) from the learned configurations, and randomly selects one as the search root. 
In the gradient descent process, \pname{} expands the search space from the root by checking all the adjacent configurations under configuration constraints (e.g., SSD capacity) in each searching 
iteration. Given the capacity constraint, \pname{} tunes one or two relevant parameters at one time, and then keep those parameters that satisfy the constraint. 

For other parameters, \pname{} will adjust their values back and forth in the search space. Once we finalize the parameters for one configuration, \pname{} will use the GPR model 
to identify the configuration with the best predicted performance grade (\circleb{4}). If its performance grade is better than the search root, \pname{} will set this configuration 
as the new search root and continue the next search iteration. 


The main challenge with the SGD procedure (\circleb{3}) is to balance the learning accuracy and exploitation overhead. 
Since there is no guarantee that the initial configuration set will cover the entire search space, \pname{} has to gradually expand its search space to ensure it can identify the 
optimal ones. However, this may cause a search space explosion. To address this issue, we introduce a heuristic exploit factor, which is the minimum Manhattan distance~\cite{manhattan} 
between the configuration being exploited and the existing learned configurations. We also set a threshold for the number of search iterations (20 iterations by default in \pname{}) 
in the configuration exploration. 

\paragraph{Predict the grades of explored configurations}
As discussed briefly in previous descriptions, \pname{} uses GPR~\cite{gp:book2006} to predict 
the grades for new configurations (\circleb{4}). This is for three major reasons. First, 
GPR can provide nearly the same performance as the deep neural networks, especially in the modeling of 
searching optimal configurations and making recommendations. Second, it offers excellent trade-offs 
between the explorations of new knowledge and learned knowledge~\cite{gp:nips2011, gp:icml2010}. 
Third, GPR provides confidence intervals with low computation overhead by default~\cite{vanaken21}.

In \pname{}, we build a new GPR model by specifying its mean function and covariance function. 
The mean function is configured as trainable, as the mean of the performance metrics 
is unknown before the learning in \pname{}. We use the covariance function to represent 
the correlation between two adjacent points in the model, and adopt both radial basis function (RBF) kernel~\cite{rbf} and rational quadratic 
kernel~\cite{rqc} as the regression covariance. We also add a white kernel~\cite{whitekernel} for random noise simulation. 


\paragraph{Validate the explored configurations}
The learning procedure of a new configuration will terminate by checking two conditions: (1) no configuration is better than the current root configuration in the search space; or (2) the 
search exceeds the threshold for the number of iterations. After that, \pname{} will use a cycle-accurate SSD simulator to validate the efficiency of the learned configurations (\circleb{5}). 
\pname{} will run all the available workloads in the \pdb{} with the SSD simulator, and report the grade for the tested configuration. 
Before the validation, \pname{} will warm up the SSD simulator by running diverse workload traces randomly. 
In the validation, \pname{} maintains a set of optimal configurations whose grades rank at the top of all the learned configurations. After a certain number of search iterations, 
if the overall grade of this configuration set is not significantly updated, the learning procedure will be converged. Otherwise, \pname{} will update the \pdb{} with the new learned configuration 
and start another search iteration until the learning procedure is converged.

\subsection{Implementation Details}
\label{subsec:impl}

We implement the \pname{} framework with Python programming language. \pname{} supports the storage traces collected with \emph{blktrace} 
which is available on a majority of computing systems. It uses Principal Component Analysis and k-means algorithms in the learning-based 
workload clustering. \pname{} utilizes the Sklearn library~\cite{www:scikit} to develop the statistic learning model that supports both SGD 
and GPR algorithms. \pname{} adopts the MQSim~\cite{mqsim:fast2018} as the backend SSD simulator to validate the learned configurations. 
Note that \pname{} is also compatible with other SSD simulators, therefore, SSD vendors can replace the open-sourced MQSim simulator with 
their own simulators. \pname{} implements the \pdb{} with the key-value store LevelDB, in which the key is the workload cluster ID, 
and the value includes the corresponding SSD configurations and their performance obtained from the SSD simulator. The value is 
organized in JSON format. \pname{} provides a simple interface \emph{set\_cons (capacity, interface, flash\_type)} to enable end users to specify 
their configuration constraints SSD capacity, interface (i.e., NVMe or SATA), and the flash type (i.e., SLC, MLC, and TLC). 
We will open source \pname{} to benefit future study. 

\subsection{Discussion and Future Work}
\label{subsec:discussion}

In this work, \pname{} mainly works under the configurations constraints that include storage cpacity, interfaces, and flash types. 
However, its learning techniques and workflow are also suitable for identifying the optimal SSD configurations with other constraints, 
such as the economic cost and energy efficiency of the SSD. Unfortunately, a majority of SSD vendors are not willing to open source 
the specifications of the cost and energy consumption of each hardware component of the SSD. Therefore, we do not study these 
configuration constraints in this work. We wish to explore these dimensions as the future work.

\section{Evaluation}
\label{sec:eval}
Our evaluation shows that: (1) \pname{} can learn optimal SSD configurations for a given workload, and the learned configurations 
can deliver improved storage performance, compared with commodity SSD configurations ($\S$\ref{subsec:learnedconfig}); 
(2) \pname{} can instantly learn an optimal configuration with low performance overheads ($\S$\ref{subsec:overhead}); 
(3) \pname{} works efficiently under different configuration constraints ($\S$\ref{subsec:constraints}); and 
(4) \pname{} itself is also tunable for satisfying various performance requirements from end users ($\S$\ref{subsec:robustness}).

\subsection{Experimental Setup}
\label{subsec:setup}

\begin{table}[t]
\scriptsize
       \caption{Application workloads used in our evaluation.}
    \label{tab:workloads} 
        \vspace{-3ex}
    \centering
\begin{tabular}{|ll|}
\hline
\multicolumn{1}{|l|}{\textbf{Workload Category}}                         & \textbf{Description}                 \\ \hline
\multicolumn{1}{|l|}{KVStore}                         & The YCSB benchmarks are executed against LevelDB.     \\ \hline
\multicolumn{1}{|l|}{Database}       & TPCC and TPCE are executed against MySQL.          \\ \hline
\multicolumn{1}{|l|}{LiveMaps}                  & The storage trace collected from a map service. \\ \hline
\multicolumn{1}{|l|}{Advertisement} & The storage trace collected from advertisement servers.  \\ \hline
\multicolumn{1}{|l|}{BatchDataAnalytics}                 & The storage trace collected from MapReduce workloads.      \\ \hline
\multicolumn{1}{|l|}{WebSearch}                       & The storage trace collected from web search services.      \\ \hline
\multicolumn{1}{|l|}{CloudStorage}                    & The storage trace collected from a cloud storage service.   \\ \hline
\end{tabular}
	\vspace{-3ex}
\end{table}

\begin{table*}[h]
\scriptsize
	\caption {Performance of the learned configurations for NVMe MLC SSDs (normalized to Intel 590 SSD).}
    \label{tab:nvmemlcperf} 
        \vspace{-3ex}
    \centering
\begin{tabular}{|l|lllllll|}
\hline
BLC/TGT         & Advertisement     & KVStore             & Database           & WebSearch          & BatchDataAnalytics    & CloudStorage       & LiveMaps            \\ \hline
Advertisement  & \textbf{3.99/1.00} & 1.67/1.00           & 1.59/0.99          & 1.50/1.00          & 1.61/1.00         & 1.62/1.00         & 1.30/1.00           \\
KVStore         & 2.13/1.00          & \textbf{31.11/1.00} & 30.96/1.00         & 6.06/1.00          & 30.67/1.00         & 30.84/1.00          & 22.41/1.00          \\
Database        & 3.20/1.50          & 6.64/1.50           & \textbf{6.89/1.50} & 3.94/1.50          & 6.84/1.50         & 6.86/1.50          & 4.11/1.49           \\
WebSearch       & 1.00/1.00          & 1.25/1.00           & 1.26/1.00          & \textbf{1.28/1.00} & 1.22/1.00          & 1.27/1.00          & 0.98/1.00           \\
BatchDataAnalytics & 1.23/1.22          & 9.48/5.25           & 9.69/5.25          & 1.98/1.17          & \textbf{ 9.81/5.25 } &  9.49/5.25          & 9.81/3.50           \\
CloudStorage    & 1.35/1.19          & 28.11/4.16           & 27.72/4.16          & 1.58/1.14          &28.82/4.16          & \textbf{29.30/4.16} & 11.11/3.57           \\
LiveMaps        & 0.67/1.00          & 2.94/1.00           & 3.07/1.00          & 4.95/1.00          &  2.99/1.00         &  3.21/1.00          & \textbf{34.61/1.00} \\ \hline
Average        & 1.94/1.13          & 11.6/2.13           & 11.60/2.12          & 3.04/1.12          &  11.70/2.13         &  11.80/2.13          & {12.04/1.79} \\ \hline
\end{tabular}
\end{table*}

In our evaluation, we use 7 different workload categories as shown in Table~\ref{tab:workloads}. These workloads cover various 
workload types that include key-value stores, databases, map services, advertisement recommendations, batch data analytics, 
web search services, and cloud storage~\cite{storagetrace:iswc2008}. Each workload type includes 
multiple storage traces. All the storage traces are either collected from university servers or enterprise servers.  


We run the \pname{} framework on a server, which is configured with 48 Intel Xeon CPU (E5-2687W v4) processors running at 3.0GHz, 96GB DRAM, and 4TB SSD. 
Since \pname{} uses the statistic learning models, it does not require GPUs in its learning procedure. We use the configurations of Intel 590 SSD, 
Samsung 850 PRO SSD and Z-SSD as the baselines, and compare the learned configurations with them to evaluate the efficiency of \pname{}. 


\subsection{Efficiency of Learned Configurations}
\label{subsec:learnedconfig}

\begin{table}[]
\scriptsize
	\caption {Learned configurations for different workloads. \footnotesize{AD: Advertisement, KV: KVStore, DB: Database, WS: WebSearch, BD: BatchDataAnalytics, CS: CloudStorage, LM: LiveMaps.}}
    \label{tab:nvmemlcconf} 
        \vspace{-3ex}
    \centering
\begin{tabular}{|l|lllll|}
\hline
Parameters  & Intel 590 & AD    & KV/DB/CS/BD & WS & LM \\ \hline
IOQueueDepth          & 8192      & 4096  & 8192  & 4096  & 8192  \\
QueueFetchSize        & 3072      & 3072  & 2048  & 4096  & 3072  \\
DataCacheCapacity     & 800MB     & 512MB & 2GB   & 800MB & 2GB   \\
CMTCapacity             & 2MB       & 2MB   & 2MB   & 4MB   & 8MB   \\
PageAllocationScheme & CWDP      & PWCD  & CDPW  & PDCW  & PWCD  \\
OverprovisioningRatio   & 0.22      & 0.22  & 0.22  & 0.20   & 0.20   \\
FlashChannelCount     & 12        & 12    & 16    & 16    & 20    \\
ChipNoPerChannel    & 5         & 6     & 4     & 5     & 7     \\
DieNoPerChip        & 8         & 8     & 8     & 4     & 2     \\
PlaneNoPerDie       & 1         & 1     & 1     & 1     & 1     \\
BlockNoPerPlane     & 512       & 512   & 512   & 1024  & 1024  \\
PageNoPerBlock      & 512       & 512   & 512   & 512   & 512   \\
PageSize            & 4096      & 4096  & 16384 & 4096  & 8192  \\ \hline
\end{tabular}
	\vspace{-3ex}
\end{table}

We first evaluate the efficiency of the learned configurations with \pname{}. We use the Intel 590 SSD as the reference. 
We set the configuration constraints as \emph{[SSD capacity = 1TB, interface = NVMe, flash type = MLC]}. 
With its configuration in the SSD simulator, we run all the workloads in Table~\ref{tab:workloads} to measure their performances. 
After that, we use the reference configuration and the measured performances to initialize \pdb{}. And then, we feed the storage 
traces from different workload types into \pname{} to learn new configurations. Once the learning converges, \pname{} will report 
the best SSD configuration for each workload. 

We show the performance of learned configurations in Table~\ref{tab:nvmemlcperf}. 
In comparison with Intel 590 SSD, the learned SSD configurations can reduces the storage latency by 1.28--34.61$\times$ for 
the target workload, while decreasing the storage latency by 8.38$\times$ on average for non-target workloads. The learned configurations can also 
improve the storage throughput by up to 5.25$\times$ for bandwidth-intensive applications such as the database and cloud storage workloads, 
without hurting the throughput of other workloads.  



To further understand the learned configurations, we list the critical parameters of the learned configurations in Table~\ref{tab:nvmemlcconf}, 
in comparison with the reference configuration of Intel 590 SSD. As we can see, for different target workloads, \pname{} will learn different values for 
these parameters, although some workload types share the same configurations. To achieve improved SSD performance, \pname{} increases the number of channels 
for most of the workloads, while adjusting the flash chip layout (e.g., the number of chips per channel) accordingly to satisfy the SSD capacity constraint. 
\pname{} will also adjusts the page allocation scheme~\cite{performeval:acmtrans} to optimize the data layout for different workloads. Similarly, \pname{} 
tunes the SSD DRAM capacity and I/O queue depth according to workload characteristics. 
Our work \pname{} proves that it is feasible to utilize ML techniques to automate the tuning of SSD configurations.  



\begin{figure}[t] 
\includegraphics[width=0.45\textwidth]{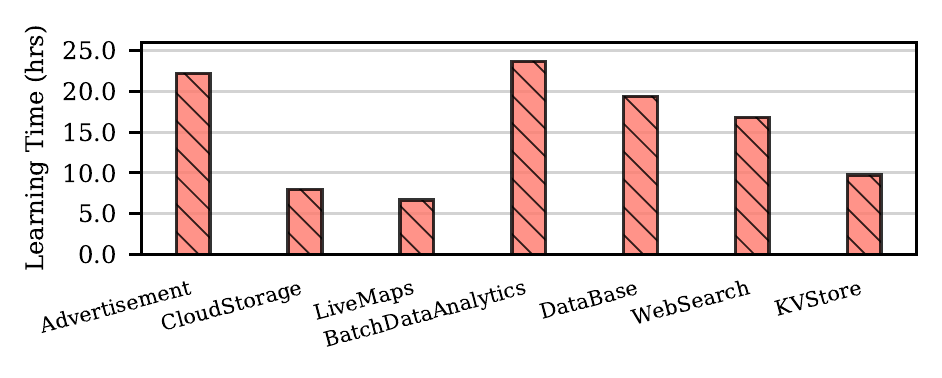}
\vspace{-4ex} 
	\caption{Learning time of \pname{} for different workloads.} 
\label{fig:training_time} 
\end{figure}

\subsection{Learning Time of \pname{}}
\label{subsec:overhead}

We now examine the learning time of \pname{}. We report the numbers for different target workloads in Figure~\ref{fig:training_time}.
\pname{} can learn an optimal configuration in 6.65--23.70 hours. And it will incur 121 search iterations on average to pinpoint the 
optimal configuration. To further understand the overhead source of \pname{}, we profile the execution time of its critical components 
on the multi-core server as described in $\S$\ref{subsec:setup}, and show the results in Table~\ref{tab:performanceoverhead}.
Our profiling results demonstrate that \pname{} can finish each search iteration within only 37.3 seconds. And the major 
performance overhead of \pname{} comes from the simulator validation, as we need to warm up the simulator before each validation. 
However, \pname{} only needs to validate the best configuration (selected based on the predicted grade with GPR) in each search iteration. 



\begin{table}[t]
\scriptsize
       \caption {Overhead sources of \pname{}.}
    \label{tab:performanceoverhead} 
        \vspace{-3ex}
    \centering
\begin{tabular}{|r|c|}
\hline
    Component                               & Execution Time (secs) \\ \hline
    Extract workload features per 100K I/O requests  & 0.84 \\ \hline
    Workload Similarity Comparison &      4.65    \\ \hline
    Workload Clustering &      0.57     \\ \hline
    \pdb{} Database Lookup             &      0.02            \\ \hline
New configuration learning per iteration &    37.3          \\ \hline
Simulator Validation                        & 691.1           \\ \hline
\end{tabular}
\vspace{-3ex}
\end{table}

\begin{table*}[]
\scriptsize
	\caption {Performance of learned configurations for NVMe SLC SSDs (normalized to Samsung Z-SSD).}
    \label{tab:nvmeslcperf} 
        \vspace{-3ex}
    \centering
\begin{tabular}{|l|lllllll|}
\hline
BLC/TGT         & Advertisement     & KVStore             & Database           & WebSearch          & BatchDataAnalytics    & CloudStorage       & LiveMaps            \\ \hline
Advertisement  & \textbf{3.99/1.00} & 3.35/1.00           & 1.21/1.00          & 3.25/1.00          & 3.64/1.00          & 3.92/1.00          & 3.71/1.00           \\
KVStore         & 2.13/1.00           & \textbf{11.49/1.00} & 2.57/1.00          & 0.78/1.00          & 9.65/1.00          & 2.17/1.00          & 9.53/1.00           \\
Database        & 3.20/1.50          & 1.31/1.49           & \textbf{3.61/1.50} & 0.91/1.50          & 3.33/1.49          & 3.26/1.50          & 2.00/1.49           \\
WebSearch       & 1.00/1.00          & 1.01/1.00           & 0.63/1.00          & \textbf{3.78/1.00} & 0.88/1.00          & 0.99/1.00          & 1.01/1.00           \\
BatchDataAnalytics & 5.18/4.01          & 1.50/2.00           & 8.72/4.01          & 11.34/4.01          & \textbf{9.72/4.00} & 5.04/4.01          & 1.59/2.00           \\
CloudStorage    & 14.17/8.00          & 1.55/1.60           & 28.04/1.60          & 14.24 / 1.60         & 21.99/1.60          & \textbf{14.25/8.00} & 2.43/2.00           \\
LiveMaps        & 0.67/1.00         & 0.44/1.00           & 6.98/1.00          & 7.5/1.00           & 1.05/1.00          & 0.66/1.00          & \textbf{19.28/1.00} \\ \hline
Average        & 4.03/2.30         & 2.95/1.30           & 7.39/1.59          & 5.97/1.59           & 7.18/1.59          & 4.33/2.50          & {5.65/1.36} \\ \hline
\end{tabular}
\end{table*}

\begin{table*}[]
\scriptsize
	\caption {Performance of learned configurations for SATA MLC SSDs (normalized to Samsung 850 PRO).}
    \label{tab:satamlcperf} 
        \vspace{-3ex}
    \centering
\begin{tabular}{|l|lllllll|}
\hline
BLC/TGT         & Advertisement     & KVStore            & Database           & WebSearch          & BatchDataAnalytics    & CloudStorage       & LiveMaps           \\ \hline
Advertisement  & \textbf{1.41/1.00 } & 1.05/1.00          & 1.40/1.00          & 1.07/1.00          & 1.39/1.00          & 1.06/1.00          & 1.02/1.00          \\
KVStore         &  1.00/1.00          & \textbf{1.01/1.00} & 0.99/1.00          & 1.00/1.00          & 1.00/1.00          & 0.97/1.00          & 0.98/1.00          \\
Database        & 1.00/1.00          & 0.99/1.00          & \textbf{1.01/1.00} & 0.95/1.00          & 1.00/1.00          & 1.04/1.00          & 1.03/1.00          \\
WebSearch       & 1.00/1.00          & 1.01/1.00          & 1.02/1.00          & \textbf{1.04/1.00} & 0.99/1.00          & 1.02/1.00          & 0.98/1.00          \\
BatchDataAnalytics & 1.03/1.00          & 0.97/1.00          & 1.00/1.00          & 1.04/1.00          & \textbf{1.07/1.00} & 1.03/1.00          & 1.05/1.00          \\
CloudStorage    & 1.56/1.31           & 2.42/2.63          & 1.64/1.31          & 2.41/2.63           & 1.22/1.11          & \textbf{2.38/2.63 } & 2.34/2.63           \\
<<<<<<< HEAD
LiveMaps        &  1.67/1.00           & 8.30/1.00          & 1.75/1.00          & 8.19/1.00          & 1.01/1.00          & 8.32/1.00         & \textbf{8.41/1.00} \\ \hline
Average        &  1.24/1.04           & 2.25/1.23          & 1.26/1.04          & 2.24/1.23          & 1.09/1.02          & 2.16/1.23         & {2.26/1.23} \\ \hline
\end{tabular}
        \vspace{-2ex}
\end{table*}

\begin{figure*}[t] 
\includegraphics[width=0.95\textwidth]{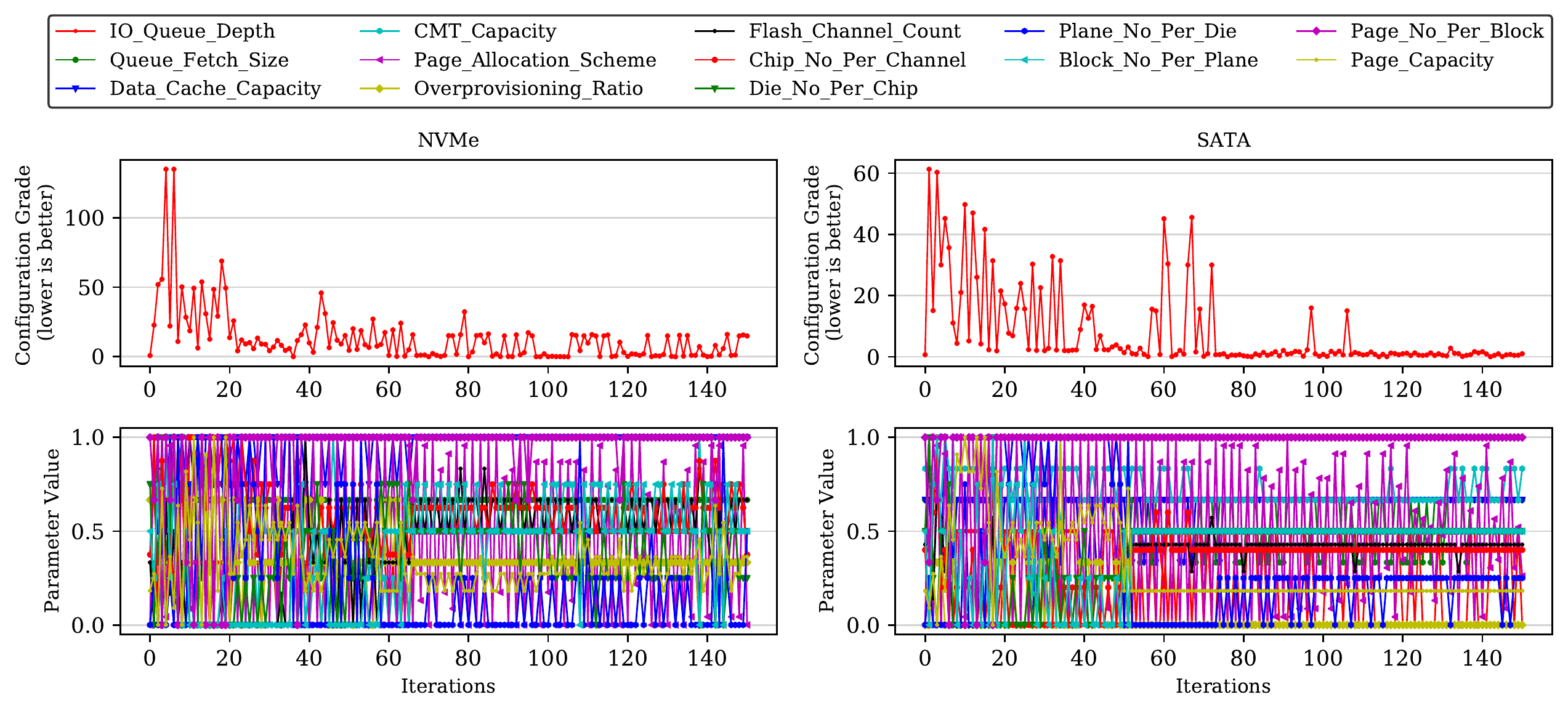}
\vspace{-3ex} 
	\caption{The learning procedure of \pname{} for NVMe and SATA SSDs, as we target batch data analytics workload.} 
\label{fig:record} 
\end{figure*}

\subsection{Sensitivity to Configuration Constraints}
\label{subsec:constraints}
We now evaluate how \pname{} performs as we change the configuration constraints that include the flash types and device interface. 
To evaluate the sensitivity to flash types, we use Samsung Z-SSD, which is a NVMe SLC SSD, as the reference configuration. 
To evaluate the sensitivity to device interface, we use Samsung 850 PRO, which is a SATA MLC SSD, as the reference configuration. 

We present the performance of the learned configurations for different workloads in Table~\ref{tab:nvmeslcperf} and Table~\ref{tab:satamlcperf} 
respectively. Table~\ref{tab:nvmeslcperf} shows that the configurations learned by \pname{} can reduce the storage 
latency by 3.61--19.28$\times$, and improve the storage throughput by up to 8.00$\times$ for NVMe SLC SSDs for the target workload, compared to the Samsung Z-SSD. 
Table~\ref{tab:satamlcperf} demonstrates that our learned configurations can deliver up to 8.41$\times$ latency reduction and 2.63$\times$ 
throughput improvement for SATA SSDs for the target workload, in comparison with Samsung 850 PRO.

In order to understand how \pname{} tunes the device parameters, we record its learning procedure and configuration grades at runtime. 
As shown in Figure~\ref{fig:record}, we present the profiling results of learning the optimal configurations for the target 
workload \emph{BatachDataAnalytics} for NVMe and SATA SSDs, respectively. The top subfigures in Figure~\ref{fig:record} demonstrate how the configuration 
grade will be updated after each search iteration. As discussed in $\S$\ref{subsec:autotuning}, 
the learning procedure will converge when the grade of the configurations becomes stable. We show the learning procedure of the critical SSD parameters 
in the below subfigures in Figure~\ref{fig:record}. We observe that, with different configuration constraints (NVMe vs. SATA), (1) the learning procedure 
will be different; (2) for each parameter, its correlation with the SSD performance is also different, making it impossible for developers to manually 
tune them; (3) not all parameters are equal, some parameters are insensitive to storage performance. \pname{} framework can help developers identify such 
parameters for different workloads under different configuration constraints, which could improve the productivity of SSD development.  

\subsection{Performance Impact of the Balance Coefficient}
\label{subsec:robustness}
\begin{figure}[t] 
\includegraphics[width=0.45\textwidth]{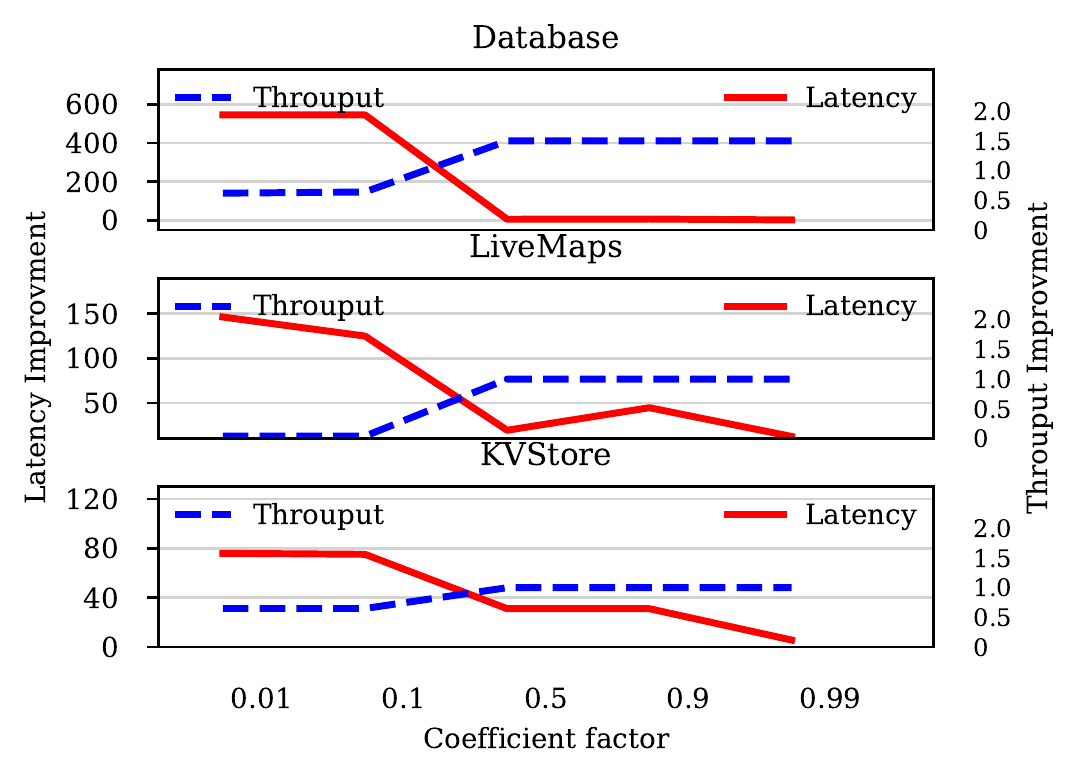}
\vspace{-3ex} 
\caption{Performance impact of the coefficient factor for balancing the latency and throughput for a target workload.} 
\label{fig:latthpt_balance} 
\end{figure}

\begin{figure}[t] 
\includegraphics[width=0.45\textwidth]{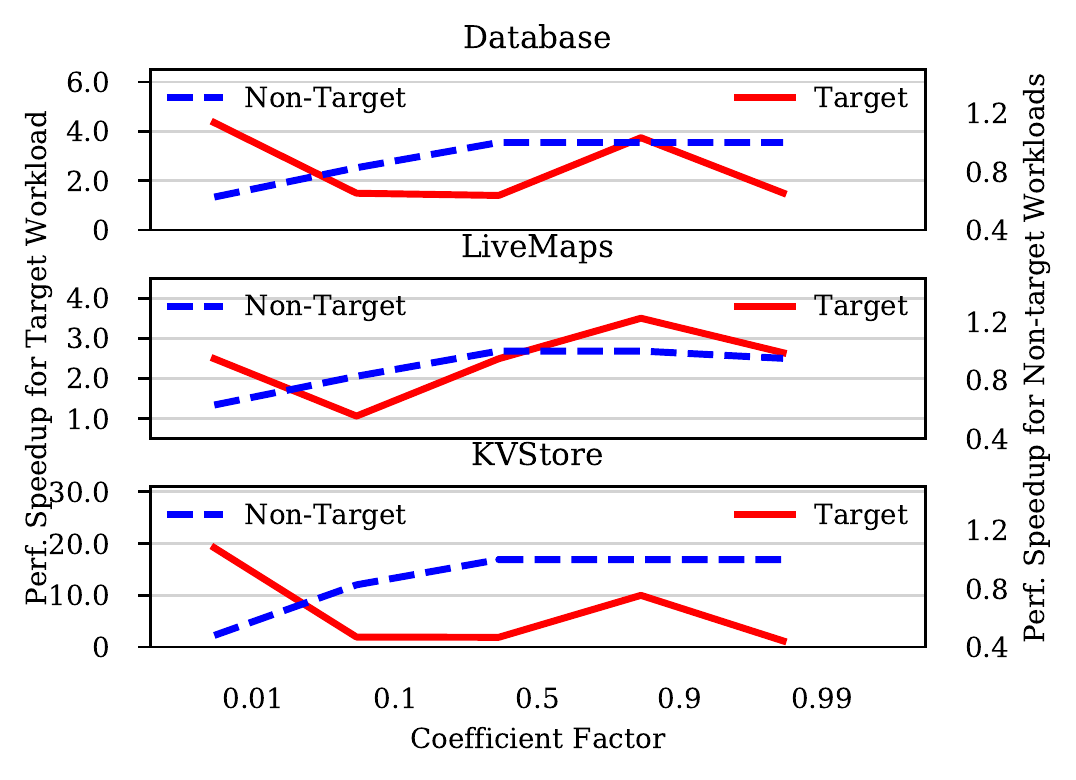}
\vspace{-3ex} 
\caption{Performance impact of the coefficient factor for balancing the performance between the target workload and non-target workloads.} 
\label{fig:penalty_balance} 
\vspace{-2ex} 
\end{figure}

As discussed in $\S$\ref{subsec:parameter} and $\S$\ref{subsec:autotuning}, \pname{} uses the coefficient factor $\alpha$ (Formula~\ref{equa:goal}) to balance 
the storage latency and throughput in the learning procedure, and defines the coefficient factor $\beta$ (Formula~\ref{equa:grade}) 
to balance the penalty (weight) between the target workload and non-target workloads. Both of them are tunable in \pname{}, which allows 
end users to adjust them per their needs. In this part, we evaluate their impact on storage performance. We vary their values from 0.01 to 0.99, and 
measure the performance of the learned configurations for the three representative workloads database, key-value store, and LiveMaps. 
As we examine each value of $\alpha$ and $\beta$, we reset the ML model and initialize the \pdb{}. 
We show the experimental results in Figure~\ref{fig:latthpt_balance} and Figure~\ref{fig:penalty_balance}.

With the coefficient factor $\alpha$, our goal is to achieve the maximum improvement for both latency and throughput. 
In Figure~\ref{fig:latthpt_balance}, as we increase the value of $\alpha$ from 0.01 to 0.3, the latency of the target workload is dramatically 
improved, however, its throughput is lower than the reference configuration. As we further increase its value to 0.9, we can achieve both improved latency 
and throughput for all the three target workloads. Thus, \pname{} sets $\alpha = 0.9$ by default.   

With the coefficient factor $\beta$, our goal is to achieve the maximum performance improvement for both the target workload and non-target workloads. 
As \pname{} learns new configurations, it is usually easy to achieve the improved performance for the target workload. However, this may decrease the 
performance for non-target workloads, which could impede the widespread adoption of the learned configurations. As we vary the value of $\beta$, 
we observe that there is such a sweet spot ($\beta = 0.9$) that can delivers maximal performance improvement for the target workload, while having 
minimal negative impact on the non-target workloads. 




\vspace{-2ex} 

\section{Related Work}
\label{sec:related}

\noindent
\textbf{SSD Performance Optimization.}
SSDs has been widely used in modern storage systems to meet the I/O performance and storage capacity requirements
of data-intensive applications, such as databases, cloud storage, web search, and big-data analytics~\cite{zheng:fast2015, kyrola:osdi2012, 
gpugraphics, tricore:sc2018, graphsurvey, liu:atc2019}. 
Although these applications have high demands on I/O performance and their workload have unique data access
patterns~\cite{iopattern:hpdc2015, deltafs:pdsw2015, iodetection:pad2019},
they normally employ generic SSD devices~\cite{ali:sc2018, metadata:ccgrid2020, flasharray:tpds2019}, which causes 
suboptimal performance and resource efficiency. 
In this paper, we develop \pname{} to facilitate the development of customized
SSD devices for applications with improved performance. 
Recently, researchers proposed the software-defined flash and open-channel SSDs to enable applications to develop 
their own storage stack for improved storage utilization and performance isolation~\cite{ouyang:asplos2014, flashblox, amf:fast2016}. 
They show that there is an increasing demand on software-defined storage. However, there is 
a longstanding gap between the application demands and device specifications. We develop \pname{} with the goal of bridging this gap.

\noindent
\textbf{Machine Learning for Systems.}
Most recently, researchers have started to leverage machine learning techniques to solve system optimization problems, 
such as the task scheduling~\cite{firm:osdi2020,mlcontainer:arxiv2021, smartharvest:eurosys2021},
cluster resource management~\cite{mao:sigcomm2019,rlscheduler:sc2020, mlcenter:2020, resource:sosp2017,
harvestvm:osdi2020}, performance optimizations~\cite{linnos:osdi2020, ionet:report2021, memoryallocation:asplos2020, zhou:mlsys2021},
data management~\cite{ottertune:sigmod2017, mldatabase:dataengr2021, vanaken:vldb2021, qtune:vldb2019},
and others~\cite{aichip:nature, apollo:arxiv2021}. However, few studies conduct a systematic investigation of applying 
the learning techniques to develop SSD devices. To the best of our knowledge, \pname{} is the first work that 
utilize the learning techniques to enable the automated tuning of SSD specifications. We believe it will not only 
benefit SSD vendors and manufacturers but also platform operators such as those for cloud services and data centers.  

\noindent
\textbf{SSD Device Development.}
Along with the architecture innovation, the industry community 
has developed mature manufacturing techniques and fabrication process to produce new storage devices,
such as Z-SSD~\cite{znand_latency}, Optane SSD~\cite{optane_latency}, ZNS SSDs~\cite{samsung:znsssd, wdc:znsssd}.  
As the industrial revolution has moved into the fourth/fifth generation (Industry 4.0/5.0) powered
by the artificial intelligence~\cite{industry4, industry5}, storage devices should also become highly customizable for
applications. Unfortunately, we are lacking an effective framework that can transfer application
demands into storage device development. In this work, we focus on building a learning-based framework 
to address a critical challenge with the SSD development -- how to efficiently identify the optimal SSD specifications for 
meeting the needs from target applications under constraints.

\section{Conclusion}
\label{sec:conclusion}
We build a learning-based framework named \pname{} for enabling the 
automated tuning of SSD specifications. Given a storage workload, 
\pname{} can efficiently learn an optimal SSD configuration that delivers the maximum 
performance improvement even under different configuration constraints. \pname{} 
can significantly reduce the manual efforts in the SSD device development. 
Our experiments show that our learned SSD configurations can significantly 
improve the storage performance for a target workload, without hurting the performance 
of non-target workloads.

%
\balance{}
\bibliographystyle{plain}
\bibliography{ref,ssdref,newref,refs}

\end{document}